# Unconventional superconductivity in highly-compressed unannealed sulphur hydride

E. F. Talantsev[1,2,*]

[1]M.N. Mikheev Institute of Metal Physics, Ural Branch, Russian Academy of Sciences, 18, S. Kovalevskoy St., Ekaterinburg, 620108, Russia

[2]NANOTECH Centre, Ural Federal University, 19 Mira St., Ekaterinburg, 620002, Russia

[*]E-mail: evgeny.talantsev@imp.uran.ru

*Abstract*

While great scientific efforts focus on the synthesis and studies of near-room-temperature (NRT) superconductors exhibited record superconducting transition temperatures (for instance, laser annealed $H_3S$, $LaH_{10}$ and $YH_n$ ($n = 4,6,7,9$) with $T_c > 200$ K), unannealed low-$T_c$ counterparts of NRT superconductors stay in the background. However, the formers are part of hydrogen-rich superconductors family and the success in understanding of NRT superconductivity depends on the study of these materials too. In this paper we analyse experimental temperature dependent upper critical field data, $B_{c2}(T)$, reported by Drozdov *et al* (*Nature* **525**, 73 (2015)) for unannealed highly-compressed ($P = 155$ GPa) sulphur hydride with $T_c = 46$ K and show that this material is unconventional superconductor which exhibits the ratio of $T_c$ to the Fermi temperature, $T_F$, in the range of $0.02 \leq T_c/T_F \leq 0.05$.

# Unconventional superconductivity in highly-compressed unannealed sulphur hydride

## I. Introduction

The discovery of near-room-temperature (NRT) superconductivity in highly compressed sulphur hydride, $H_3S$, by Drozdov *et al.* [1] and in lanthanum decahydride, $LaH_{10}$, by Somayazulu *et al.* [2], raises a wide public interest and great scientific efforts to study pressure-induced superconductivity in hydrogen-rich materials [3-7]. It should be noted, that the discovery of the effect of pressure-induced superconductivity in non-superconducting materials is attributed to Jörg Wittig [8] who converted elemental cerium into superconductor at pressure of $P = 5$ GPa. During this, more than fifty years, long research journey, pressure-induced superconductivity was found in dozens of non-superconducting elements and compounds (details can be found in recent reviews [3,9-11]).

It is important to note, that all known NRT superconductors (which exhibits record superconducting transition temperatures, $T_c$) are synthesized by the laser annealing technique of highly-compressed hydrogen-rich precursors in the diamond-anvil cell (DAC) [1,2,10-15]. At the same time, some unannealed phases, and particularly unannealed sulphur hydride, which alternatively can be $H_2S$ [16,17], $(H_2S)_2H_2$ [18], or $H_3S$ [1], are also superconductors with, however, much lower $T_c$. Direct experimental studies reported by Einaga *et al.* [19] showed that NRT superconducting phase of laser annealed sulphur hydride at pressure in the range of 150 GPa $\leq P \leq$ 225 GPa has *Im-3m* symmetry, while lower pressure phase ($P \leq 150$ GPa) has *R3m* symmetry. This result is significantly different from theoretically predicted pressure range of 130 GPa $\leq P \leq$ 200 GPa [20] for thermodynamic stability of *R3m*-phase. Based on this, mentioned above theoretically possible phases of $H_2S$ [16,17] and $(H_2S)_2H_2$ [18] (for low-temperature superconducting phase of unannealed sulphur hydride) require direct experimental studies to be confirmed/disproved. Taking in account that low $T_c$ in unannealed sulphur hydride can be originated from low atomic order in $H_3S$-*Im-3m* phase (because this phase synthesizes at room temperature), we will designate studied phase as unannealed sulphur hydride, herein.

Relatively low $T_c$ in unannealed highly-compressed hydrides is the reason why these superconducting compounds are in the background of its NRT counterparts. Independent of its lower $T_c$'s, these phases are highly-compressed hydrogen-rich superconductors and there is an interest to understand the superconductivity in these compounds too.

In this paper we analyse temperature dependent upper critical field data, $B_{c2}(T)$, for unannealed sulphur hydride ($P$ = 155 GPa) reported by Drozdov *et al.* in their first milestone paper [1]. In the result, it is found that unannealed sulphur hydride has the ratio of $T_c$ to the Fermi temperature, $T_F$, in the range of $0.02 \leq T_c/T_F \leq 0.05$, and, thus, this superconductor falls in the unconventional superconductor band of the Uemura plot [21-23] together with heavy fermions, fullerens, cuprates, pnictides and NRT superconductors.

## II. Extrapolative models for ground state upper critical field

Ground state upper critical field, $B_{c2}(0)$, is given in the Ginzburg-Landau theory [24] by following expression:

$$B_{c2}\left(\frac{T}{T_c} = 0\right) = \frac{\phi_0}{2\cdot\pi\cdot\xi^2(0)}, \tag{1}$$

where $\phi_0 = 2.068 \cdot 10^{-15}$ Wb is magnetic flux quantum, and $\xi(0)$ is the ground state coherence length. To deduce $\xi(0)$ value from real world measurements (which very often perform at high reduced temperatures) several models were proposed. In this paper, to deduce $\xi(0)$ value for unannealed $H_3S$ phase we will use classical two-fluid Gorter-Casimir model (GC model) [25]:

$$B_{c2}(T) = B_{c2}(0) \cdot \left(1 - \left(\frac{T}{T_c}\right)^2\right) = \frac{\phi_0}{2\cdot\pi\cdot\xi^2(0)} \cdot \left(1 - \left(\frac{T}{T_c}\right)^2\right) \tag{2}$$

as well as model proposed by Gor'kov [26] (Gor'kov model):

$$B_{c2}(T) = B_{c2}(0) \cdot \left( \frac{1.77 - 0.43 \cdot \left(\frac{T}{T_c}\right)^2 + 0.07 \cdot \left(\frac{T}{T_c}\right)^4}{1.77} \right) \cdot \left[ 1 - \left(\frac{T}{T_c}\right)^2 \right] =$$

$$\frac{\phi_0}{2 \cdot \pi \cdot \xi^2(0)} \cdot \left( \frac{1.77 - 0.43 \cdot \left(\frac{T}{T_c}\right)^2 + 0.07 \cdot \left(\frac{T}{T_c}\right)^4}{1.77} \right) \cdot \left[ 1 - \left(\frac{T}{T_c}\right)^2 \right] \quad (3)$$

and modified Werthamer, Helfand, and Hohenberg (WHH) [27,28] model proposed by Baumgartner *et al* [29] (B-WHH model):

$$B_{c2}(T) = B_{c2}(0) \cdot \left( \frac{\left(1 - \frac{T}{T_c}\right) - 0.153 \cdot \left(1 - \frac{T}{T_c}\right)^2 - 0.152 \cdot \left(1 - \frac{T}{T_c}\right)^4}{0.693} \right) =$$

$$= \frac{\phi_0}{2 \cdot \pi \cdot \xi^2(0)} \cdot \left( \frac{\left(1 - \frac{T}{T_c}\right) - 0.153 \cdot \left(1 - \frac{T}{T_c}\right)^2 - 0.152 \cdot \left(1 - \frac{T}{T_c}\right)^4}{0.693} \right) \quad (4)$$

### III. $B_{c2}(T)$ analysis for unannealed sulphur hydride ($P = 155$ GPa)

Drozdov *et al* [1] in their Figure 3(a) reported the temperature dependent magnetoresistance, $R(T,B)$, for unannealed sulphur hydride sample pressurised at $P = 155$ GPa. To deduce raw $B_{c2}(T)$ data we processed $R(T,B)$ data (Figure 3(a), upper insert [1]) by utilising a criterion of $R(T) = 345$ mΩ. Raw $B_{c2}(T)$ data and results of fits to three models (Eqs. 2-4) are shown in Fig. 1 and Table 1.

**Table I.** Deduced and calculated parameters for unannealed sulphur hydride compressed at $P = 155$ GPa. Deduced critical temperature for all models (Eqs. 2-4) is $T_c = 45.9 \pm 0.1$ K. Assumed charge effective mass is $m^*_{eff} = 2.76 \cdot m_e$ [31]. Smallest and largest values for $\frac{T_c}{T_F}$, $\frac{T_c}{T_{fluc,phase}}$ and $\frac{T_c}{T_{fluc,amp}}$ are marked in bold.

| Model | Deduced $\xi(0)$ (nm) | Assumed $\frac{2 \cdot \Delta(0)}{k_B \cdot T_c}$ | $T_F$ ($10^3$ K) | $T_c/T_F$ | Assumed $\kappa$ | $T_{fluc,phase}$ ($10^3$ K) | $T_{fluc,amp}$ ($10^3$ K) | $T_c/T_{fluc,phase}$ | $T_c/T_{fluc,amp}$ |
|---|---|---|---|---|---|---|---|---|---|
| GC | 3.6 ± 0.1 | 3.53 | 1.3 ± 0.1 | 0.035 ± 0.002 | 60 | 1.88 ± 0.05 | 0.50 ± .02 | 0.024 ± 0.001 | 0.091 ± 0.003 |
|  |  | 4.7 | 2.3 ± 0.1 | **0.020 ± 0.01** | 120 | 0.47 ± 0.01 | 0.13 ± 0.01 | **0.098 ± 0.03** | **0.36 ± 0.01** |
| Gor'kov | 3.3 ± 0.1 | 3.53 | 1.1 ± 0.1 | 0.042 ± 0.002 | 60 | 2.05 ± 0.06 | 0.55 ± 0.02 | 0.022 ± 0.001 | 0.083 ± 0.003 |
|  |  | 4.7 | 1.95 ± 0.11 | 0.024 ± 0.001 | 120 | 0.51 ± 0.02 | 0.14 ± 0.01 | 0.090 ± 0.003 | 0.33 ± 0.01 |
| B-WHH | 3.1 ± 0.1 | 3.53 | 0.97 ± 0.06 | **0.047 ± 0.004** | 60 | 2.18 ± 0.07 | 0.59 ± 0.02 | **0.021 ± 0.001** | **0.078 ± 0.002** |
|  |  | 4.7 | 1.7 ± 0.1 | 0.027 ± 0.002 | 120 | 0.54 ± 0.02 | 0.15 ± 0.01 | 0.084 ± 0.01 | 0.31 ± 0.01 |

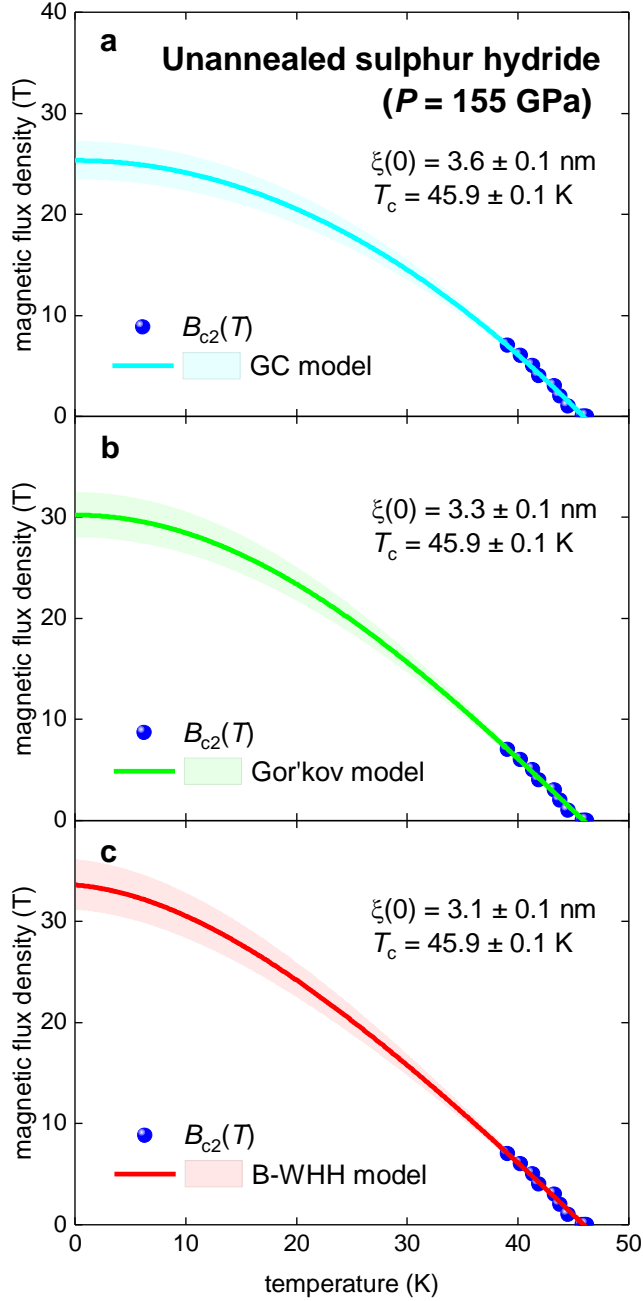

**Figure 1.** Superconducting upper critical field, $B_{c2}(T)$, data and fits to three models (Eqs. 2-4) for unannealed sulphur hydride phase compressed at pressure of $P = 155$ GPa (raw data is from Ref. 1). (a) fit to GC model, the fit quality is $R = 0.989$. (b) fit to Gor'kov model, $R = 0.991$. (c) fit to B-WHH models, $R = 0.991$. 95% confidence bars are shown.

## IV. Unannealed sulphur hydride ($P = 155$ GPa) in Uemura plot

From deduced $\xi(0)$ values (Fig. 1 and Table 1), we calculated the Fermi temperature, $T_F$, for unannealed sulphur hydride phase by utilising the Bardeen-Cooper-Schrieffer theory expression [30]:

$$T_F = \frac{\varepsilon_F}{k_B} = \frac{\pi^2}{8 \cdot k_B} \cdot m_{eff}^* \cdot \xi^2(0) \cdot \left(\frac{\alpha \cdot k_B \cdot T_c}{\hbar}\right)^2, \tag{5}$$

where $\alpha = \frac{2 \cdot \Delta(0)}{k_B \cdot T_c}$, $\Delta(0)$ is the amplitude of the ground state energy gap, $\varepsilon_F$ is the Fermi energy, $\hbar = h/2\pi$ is reduced Planck constant, $k_B$ is the Boltzmann constant, $m_{eff}^*$ is the charge carrier effective mass ($m_{eff}^* = 2.76 \cdot m_e$ for H3S at $P$ = 155 GPa [31]).

For H3S phase $\alpha$ = 3.53-4.7, where the lower bound was reported by Kaplan and Imry [32] and in our previous works [33-35], and the upper bond was reported by other authors [3,30,36-41]. Due to for all three models deduced $T_c$ = 45.9 ± 0.1 K in Table I we show only the $T_c/T_F$ ratios.

As the result, unannealed sulphur hydride ($P$ = 155 GPa) phase in all considered scenarios (Table 1) has $0.02 \leq T_c/T_F \leq 0.05$ and falls in unconventional superconductors band of the Uemura plot [21-23] in close proximity to other NRT counterparts [1,2,12-15] (Fig. 2), including annealed H3S with transition temperature $T_c$ ~ 190 K [1,42]. Two characteristic lines for Bose-Einstein condensates (BEC) (for which $T_c/T_F$ = 0.22) and the Bardeen-Cooper-Schrieffer (BCS) superconductors (for which $T_c/T_F \leq 0.001$) are shown in Fig. 2 for clarity.

Most of NRT superconductors have severe influence of thermodynamic fluctuations on the superconducting order parameter and, thus, on the observed superconducting transition temperature, $T_c$ [43-45]. There are two types of the thermodynamic fluctuations in superconductors, the phase fluctuations of the order parameter, which have characteristic temperature [46]:

$$T_{fluc,phase} = \frac{0.55 \cdot \phi_0^2}{\pi^{3/2} \cdot \mu_0 \cdot k_B} \cdot \frac{1}{\kappa^2 \cdot \xi(0)} \tag{6}$$

where $\kappa = \lambda(0)/\xi(0)$ is Ginzburg-Landau parameter, and $\lambda(0)$ is the ground state London penetration depth; and the amplitude fluctuations of the order parameter [47], with characteristic temperature of:

$$T_{fluc,amp} = \frac{\phi_0^2}{12 \cdot \pi \cdot \mu_0 \cdot k_B} \cdot \frac{1}{\kappa^2 \cdot \xi(0)} \tag{7}$$

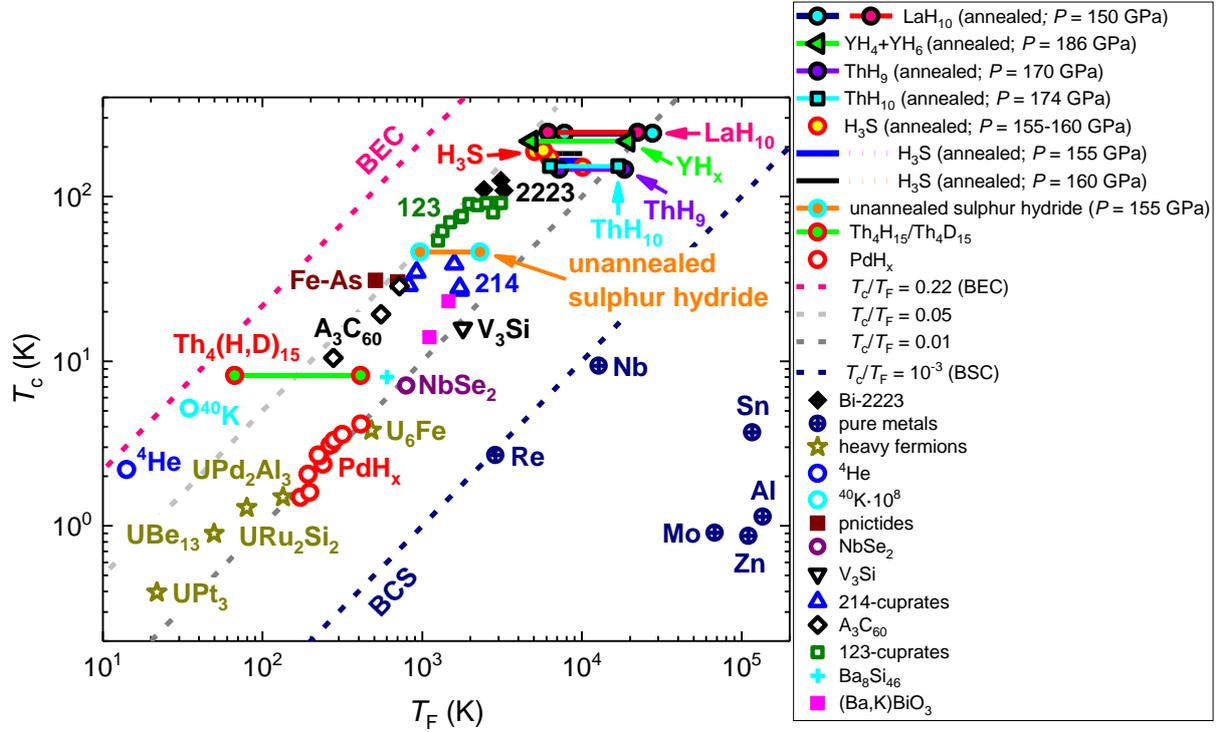

**Figure 2.** A plot of $T_c$ versus $T_F$ where the most representative superconducting families and unannealed highly-compressed sulphur hydride ($P = 155$ GPa) are shown. Other data is taken from [23,35,43-45]. Characteristic lines for the Bose-Einstein condensate (BEC) and the Bardeen-Cooper-Schrieffer (BCS) superconductors are shown for clarity.

To calculate $T_{fluc,phase}$ and $T_{fluc,amp}$ we make an assumption that unannealed $H_3S$ phase has the Ginzburg-Landau parameter in the range of $\kappa = 60-120$ which covers all superconductors with transition temperature $T_c \geq 20$ K [35,48-51].

Examination of obtained values for $T_{fluc,phase}$ and $T_{fluc,amp}$ (Table I) leaded us to an important conclusion that thermodynamic fluctuations are not influenced on the observed $T_c$ in this compound, which is different from the case of annealed $H_3S$ phase where the ratio of $T_c/T_{fluc,amp}$ in some scenarios can be as high as $T_c/T_{fluc,amp} = 0.7$ [43].

It should be noted that first-principles calculations studies [16-18,20,30,36-41] showed that experimentally observed $T_c$ in laser-annealed highly-compressed H$_3$S can be accurately computed within electron-phonon phenomenology by accepting four assumptions:

1. strong electron-phonon coupling strength, $\lambda_{e-ph} \sim 1.8$;
2. the dominance of anharmonic vibrations in the coupling;
3. reasonably high Coulomb pseudopotential $\mu^* \sim 0.18$ vs conventionally used value of $\mu^* = 0.10$.
4. and as a consequence of 1-3 above, a large value for the ratio:

$$\alpha = \frac{2 \cdot \Delta(0)}{k_B \cdot T_c} = 4.5 - 4.9 \qquad (8)$$

These results [16-18,20,30,36-41], however, do not prohibit a possibility that alternative approaches based on different coupling mechanisms can be also successful in deducing observed $T_c$ for some (perhaps, also reasonably unusual) parameters values. Alternative mechanisms for superconductivity are in discussion for several decades [52-54]. In this regard, it should be mentioned that the analysis of experimental data beyond $T_c$, for instance, the temperature dependent self-field critical current density, $J_c(\text{sf},T)$ [33], and the upper critical field, $B_{c2}(T)$ [35], in laser-annealed H$_3$S samples showed that:

$$\alpha = \frac{2 \cdot \Delta(0)}{k_B \cdot T_c} = 3.55 \pm 0.31 \qquad (9)$$

This (deduced from experiment) α is remarkably different from the computed value (Eq. 8) (calculated based on electron-phonon pairing mechanism) and, at the same time, it is so close to the BCS weak-coupling limit, that it is difficult to believe that there is a coincident. Another important issue is the isotope effect in superhydride/superdeuteride systems. Despite a fact that practically all experimentally reported $T_c$ values for laser-annealed highly-compressed sulphur deuterides are lower than ones reported for sulphur hydrides, it should be noted that the synthesis of superhydrides/superdeuterides requires extreme conditions, at which hydrogen and deuterium unlikely have the same catalytic activities. Truly, as it is shown in recent report by Drozdov *et al* [14], lanthanum superhydride and lanthanum superdeiteride have different phase stoichiometry, i.e. $LaH_{10}$ and $LaD_{11}$ respectively. And thus, there is no final clarity for the origin of the $T_c$ differences in LaH-LaD system, because the latter can manifest the stoichiometry or phases differences for superhydride and superdeiteride compounds. Also, it can be seen in Fig 3,c (in report by Einaga *et al.* [20]), that $H_3S$ and $D_3S$ have a large difference (> 10 GPa) in the position of the phase boundary between *Im-3m* and *R3m* phases. This alludes on the difference in HS-DS phases stoichiometry for laser-annealed samples. Thus, more experimental and theoretical studies on the isotope effect and phase compositions in superhydrides/superdeuterides are required.

**V. Conclusions**

In summary, in this paper we analyse experimental $B_{c2}(T)$ data of unannealed highly-compressed $H_3S$ compound and find that this superconductor exhibits unconventional superconductivity. In addition, we show that thermodynamic fluctuations do not affect the observed critical temperature $T_c$ = 46 K in this superconductor.

**Acknowledgement**

Author thanks financial support provided by the state assignment of Minobrnauki of Russia (theme "Pressure" No. AAAA-A18-118020190104-3) and by Act 211 Government of the Russian Federation, contract No. 02.A03.21.0006.